\documentclass[conference,fleqn]{IEEEtran}
\usepackage{color}

\usepackage{verbatim}
\usepackage{graphicx}        
\DeclareGraphicsExtensions{.jpg, .pdf, .mps, .png}

\newcommand{\art}[1]{\textcolor{blue}{\bf [Artur]: #1}}

\ifCLASSINFOpdf
\else
\fi
\usepackage{amssymb}
\usepackage[cmex10]{amsmath}
\usepackage[tight,footnotesize]{subfigure}
\usepackage{url}

\usepackage{multirow}

\usepackage{cite}

\usepackage{algorithm}
\usepackage{algorithmic}
\floatname{algorithm}{Procedure}

\begin{document}
%






\title{Distributed Location of the Critical Nodes to Network Robustness based on Spectral Analysis}

\author{\IEEEauthorblockN{Klaus Wehmuth and Artur Ziviani}
\IEEEauthorblockA{National Laboratory for Scientific Computing (LNCC)\\
Av, Get\'{u}lio Vargas, 333 -- 25651-075 -- Petr\'{o}polis, RJ, Brazil\\
Email: \{klaus,ziviani\}@lncc.br}
}


%



\maketitle

\begin{abstract}
We propose a methodology to locate the most critical nodes to network robustness in a fully distributed way. 
Such critical nodes may be thought of as those most related to the notion of network centrality.
Our proposal relies only
on a localized spectral analysis of a limited neighborhood around each node in the network. 
We also present a procedure allowing the navigation from any node
towards a critical node following only local information computed by the proposed algorithm. Experimental
results confirm the effectiveness of our proposal considering networks of different scales and topological characteristics.
\end{abstract}



%
\IEEEpeerreviewmaketitle

\section{Introduction}
\label{sec:Int}

Currently, very large technological networks, such as
the Internet, overlay networks, and
online social networks, play a key role in the modern
society. From the network management standpoint, it is of
upmost importance to have means to understand in an efficient way 
the level of connectivity of these networks. This directly relates to
how robust these networks are to possible partition as such knowledge 
is useful to assess potential security threats and performance bottlenecks.

The study of network robustness receives a lot of attention in many areas related to network science~\cite{Lewis2009a,Kocarev2010}, in particular
research in complex communication networks~\cite{Albert2000a,Newman2000,Cohen2000,Kim2004,Mihail2006a,Xiao2008,Yan2010}.
Network robustness basically 
relates to the analysis of topological properties of complex networks to evaluate how well such networks are connected and 
how close they are to be fragmented, thus disrupting their functionality. Although many previous works evaluate network robustness
in general~\cite{Albert2000a,Newman2000,Cohen2000,Kim2004,Mihail2006a,Xiao2008,Yan2010}, only fewer recent studies~\cite{Nanda2008,Kermarrec2011,Dinh2010} address the particular topic of identifying the most critical nodes to network robustness in a distributed way. 
Such critical nodes may be thought of as those most related to the notion of network centrality, \textit{i.e.}, the nodes presenting 
the highest impact on connectivity in the case of imminent network fragmentation or those
the most important to efficient information spreading in diffusion networks.

In this paper, we propose a distributed methodology to assess and locate critical nodes to network robustness based on
spectral analysis~\cite{Mohar1991,Chung1997,Spielman2010}~(see Section~\ref{sec:Bkg} for a background). We also present a complementary procedure that allows one to navigate from any node towards
a critical node following only local information computed by the proposed methodology. 
We evaluate the proposed methodology in different networks, ranging from synthetic generated 
networks to a real-world network trace. Results confirm the effectiveness of the proposed
methodology in locating the most critical nodes to network robustness in a distributed manner
within networks with different characteristics and scales. 

This paper is organized as follows. In Section~\ref{sec:Bkg}, we review some theoretical concepts upon which we build our proposal. 
We introduce our proposed methodology and navigation procedure in Section~\ref{sec:Des}. Experimental results are presented in Section~\ref{sec:Res}. In Section~\ref{sec:RelW}, we
analyze related work. Finally, in Section~\ref{sec:Conc}, we conclude and discuss future work.

\section{Background on Spectral Analysis}
\label{sec:Bkg}

In this section, we provide some basic definitions and  background concerning spectral analysis 
that are needed for the development of our work.

\subsection{Graph}
\label{subsec:Grp}
Consider a $n$ node network represented as an undirected graph $\mathcal{G} = (V,E)$,
with $|V|=n$ vertices and $|E|$ edges. For a node $i \in V$ we denote as $d_i$ the degree
of node~$i$. This definition is sufficient for representing undirected graphs with at most 
one connection between any given pair of different vertices and no connection from a vertex 
to itself.

\subsection{Adjacency Matrix}
\label{subsec:AdjMatrix}
The adjacency matrix $\mathbf{A}(\mathcal{G})$ of the graph $\mathcal{G}$ is defined as

\begin{equation}
\label{ADM}
\begin{split}
	\mathbf{A}_{ij} =&
	\begin{cases}
		1, & (i,j) \in E,\\
		0, & \text{otherwise.}
	\end{cases}\\
\end{split}
\end{equation}
Since the adjacency matrix is square and symmetric, it follows that it has a full set of real 
eigenvalues and orthogonal eigenvectors and can be decomposed as \mbox{$ \mathbf{A} = Q \Lambda Q^T $}, 
where $\Lambda$ is the diagonal matrix of (real) eigenvalues and $Q$ is the orthogonal matrix of eigenvectors. 
Further, it should be noticed that adding all the entries of a row on the adjacency matrix results 
in the degree of the node represented by that row. This same property also holds for adding up the entries of a column.

\subsection{Diagonal Degree Matrix}
\label{subsec:DDgMatrix}
The diagonal degree matrix $\mathbf{D}$ is the matrix that has the degree of the nodes at the diagonal and zeros elsewhere. It is 
defined as follows:
\begin{equation}
\label{DDM}
 \mathbf{D}_{ij} = 
 \begin{cases}
   d_n & \text{if } i=j=n, \\
   0 & otherwise,\\
 \end{cases}
\end{equation}
where $d_n$ is the degree of node $n$.

\subsection{Laplacian Matrix}
\label{subsec:Lap}
The Laplacian matrix of the graph $\mathcal{G}$, denoted $\mathbf{L}(\mathcal{G})$, is
defined as

\begin{equation}
\label{LPM}
\begin{split}
	\mathbf{L}_{ij} =&
	\begin{cases}
		d_i, & i = j,\\
		-1,  & (i,j) \in E,\\
		0,   & \text{otherwise.}
	\end{cases}\\
\end{split}
\end{equation}

The Laplacian of a graph is also often defined in terms of the adjacency matrix 
$\mathbf{A}$~(Eq.~\eqref{ADM}) as $\mathbf{L} = \mathbf{D} - \mathbf{A}$,
where $\mathbf{D}$ is the diagonal degree matrix (Eq.~\eqref{DDM}). It should be noted 
that the vertex sequence on the representation of $\mathbf{D}$ and $\mathbf{A}$ needs 
to be the same.
 
From the definition~\eqref{LPM} it can be seen that the sum of all columns of $\mathbf{L}$ 
is the zero vector ($\vec{0}$), as for each line there is an entry with the degree of the node 
and a $-1$ for each connection on the respective node. Therefore, it follows that $\mathbf{L}$ is 
singular and that the all ones vector ($\vec{1}$) is in the \textit{Nullspace} of $\mathbf{L}$.
As a consequence, 0 is an eigenvalue of $\mathbf{L}$ and $\vec{1}$ in an eigenvector 
associated to the eigenvalue 0.

The Laplacian matrix is also known as Combinational Laplacian matrix, and we will from now on refer to it 
this way in order to distinguish it from the Normalized Laplacian matrix defined in the following.

\subsection{Normalized Laplacian Matrix}
\label{subsec:NormLap}
The normalized Laplacian matrix~\cite{Chung1996} of the graph $\mathcal{G}$ 
is defined as $\mathcal{L}_{ij}(\mathcal{G}) =
\mathbf{I} - \mathbf{D}^{-1/2}\mathbf{A}\mathbf{D}^{-1/2}$,
where $\mathbf{I}$ is the identity matrix and $\mathbf{D}$ is the diagonal degree matrix \eqref{DDM}, that is

\begin{equation}
\begin{split}
	\mathcal{L}_{ij}(\mathcal{G}) = &
	\begin{cases}
		1, & i = j,\\
		-\frac{1}{\sqrt{d_id_j}},  & (i,j) \in E,\\
		0,   & \text{otherwise.}
	\end{cases}\\
\end{split}
\end{equation}

Alternatively the Normalized Laplacian matrix can also be defined as 
$\mathcal{L}_{ij}(\mathcal{G}) = \mathbf{D}^{-1/2}\mathbf{L}\mathbf{D}^{-1/2}$ 
and is thus closely related to the Combinational Laplacian matrix.
Nevertheless, different from the Combinational Laplacian, the Normalized Laplacian matrix $\mathcal{L}$ 
has some key properties that are of particular interest in this work~\cite{Mohar1991,Chung1997,Spielman2010}:
(i)~all its eigenvalues are between 0 and~2, \textit{i.e.}, $0 = \lambda_1(\mathcal{L}) \leq \lambda_2(\mathcal{L}) \leq \cdots \leq 
\lambda_n(\mathcal{L}) \leq 2$; 
and (ii) for networks with a single connected component, ~$\lambda_2(\mathcal{L})$ is the 
smallest non-zero eigenvalue and is less than 1 if the graph is not complete, reflecting the graph 
connectivity level approaching 0 as the graph tends to be less connected.  
This particular eigenvalue, $\lambda_2(\mathcal{L})$, also known as the \emph{spectral gap}, is extensively 
used in this work and it will be referred to simply as $\lambda_2$ hereafter. 
The property of having all eigenvalues normalized between 0 and 2 makes the normalized Laplacian matrix well suited for comparing the
spectrum of graphs with different sizes.

\section{Proposed Methodology}
\label{sec:Des}

Our goal is a distributed methodology capable of locating the critical node(s) to network robustness.
The rationale behind our proposal is that if a particular topological characteristic of the network causes the spectral gap~$\lambda_2$ to be 
low, this same characteristic also causes $\lambda_2$ to be \emph{locally} low in a relatively small neighborhood 
located around such characteristic. This brings up the concept of assigning a local value for each node based on $\lambda_2$ computed for a 
neighborhood of a given number $h$ of hops around this node. By doing so in parallel for each node in the network,
every node then has itself attributed with a local value that can be compared in
order to rank the nodes in terms of their relative local importance to network robustness.

\subsection{Locating the Critical Node(s) of the Network}
\label{subsec:cri}

We observed experimentally that the locally computed $\lambda_2$ has a bias towards higher values on nodes with higher degrees, 
thus causing the local values in principle to be over sensitive to the presence of high degree nodes.
In order to mitigate this unsuitable effect, the local value $\kappa_v$ we assign to each node $v$ is actually given by

\begin{equation}
\begin{split}
	\kappa_v =&
	\begin{cases}
		\dfrac{\lambda^v_2}{\log_2(d_v)}, & d_v > 1,\\
		\infty, & d_v = 1,
	\end{cases}\\
\end{split}
\end{equation}

\noindent
where $\lambda^v_2$ is the spectral gap of the h-neighborhood of node~$v$ (\textit{i.e.},
the subnetwork composed by all nodes within $h$ hops of node $v$),
and $d_v$ 
is the degree of node~$v$. 
If $d_v = 1$ (\textit{i.e.}, node $v$ is a leaf), then $\log_2(d_v) = 0$ and thus we consider $\kappa_v = \infty$
since a leaf is indeed the least critical node in the network.
The same radius is used to define the h-neighborhood around every node.
From this definition, we remark that each node $v$ only requires the knowledge of local information
concerning a \mbox{h-neighborhood} surrounding itself  to compute its $\kappa_v$.
Therefore, there is no need for the full network
topology to be known by any particular node and $\kappa_v$ can be computed in a fully distributed way for all
nodes within the network.

Once each node $v$ has its assigned $\kappa_v$ value, they compare it to the corresponding values of all nodes
in the h-neighborhood used to compute $\kappa_v$ and identify the node with the lowest $\kappa_v$ value. After identifying such a node,
they indicate to that node that it has the lowest $\kappa_v$ visible to them. Each node will in turn account
the indications received and use it to calculate a score $S_v$ defined as

\begin{equation}
S_v = \frac{NumberOfIndications}{|h\text{-}neighborhood_v|},
\end{equation}

\noindent
where $NumberOfIndications$ is the total number of indications received by node $v$ and $|h\text{-}neighborhood_v|$ is the number of nodes in its own 
h-neighborhood.
It follows from this definition that $0 \leq S_v \leq 1$ at each node $v$ because $NumberOfIndications$ can vary from $0$ to the number
of nodes within its h-neighborhood~(\textit{i.e.}, $|h\text{-}neighborhood_v|$).

\long\def\symbolfootnote[#1]#2{\begingroup%
\def\thefootnote{\fnsymbol{footnote}}\footnote[#1]{#2}\endgroup} 

Each node $v$ then has its own $S_v$ score; and 
at least one node in the whole network has a score $S_v = 1$.\symbolfootnote[2]{There might be rare scenarios where
no node in the  network has $S_v = 1$ if the network is regular enough to have many h-neighborhoods that are iso-spectral, as
for instance a ring, thus yielding equal lowest values to $\kappa_v$ in different nodes composing the h-neighborhood.
In these atypical cases, each node considers the node with the lowest id number as
the one with the lowest $\kappa_v$ it sees, thus ensuring at least one node has a score $S_v = 1$.}
This happens because in general there is  a minimum $\kappa_v$ value for the entire network and the node $v$ associated to it is 
thus identified as the lowest $\kappa_v$ by all of the
elements of its \mbox{h-neighborhood}. The nodes with $S_v = 1$ are defined as the \emph{critical nodes} of the network, \textit{i.e.}
the nodes that represent the most fragile points of the network robustness.

\begin{figure*}[ht]
 \centering
 \includegraphics[width=12cm,keepaspectratio=true]{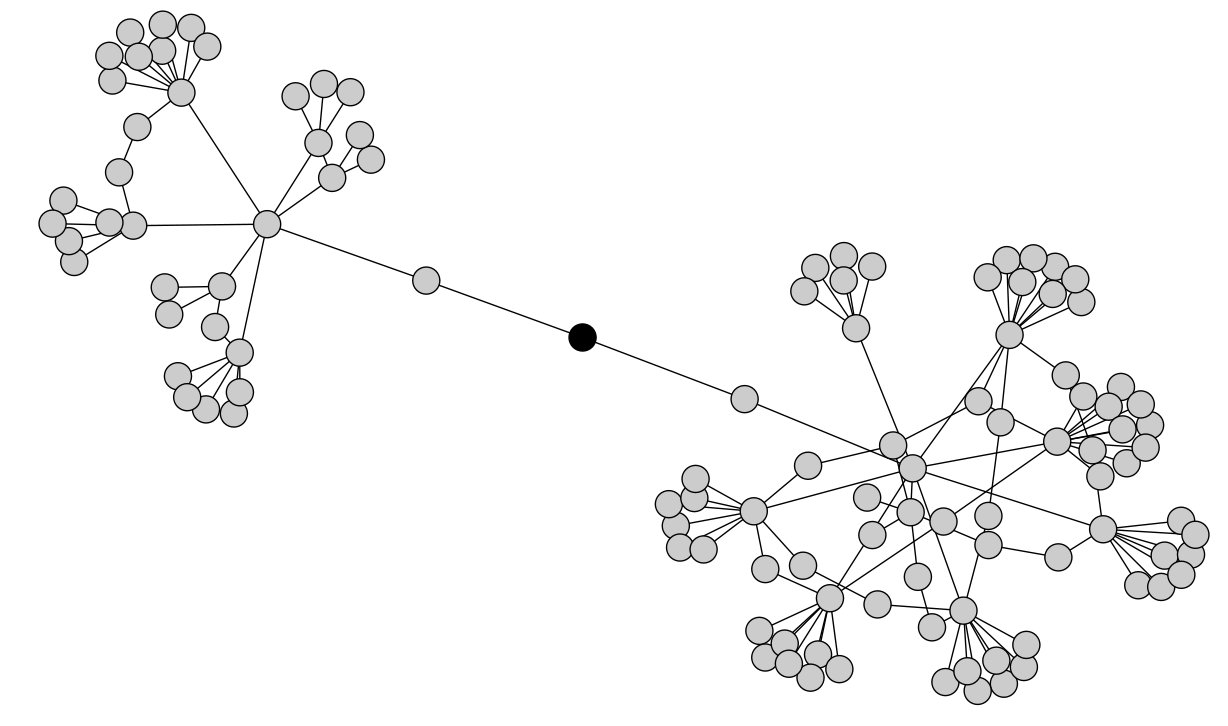}
 \caption{h-neighborhood in which a node (in black) has $S_v = 1$.}
 \label{fig:subn1}
\end{figure*}

Figure~\ref{fig:subn1} shows an example h-neighborhood established around the node in black with 4-hops around it.
In this particular h-neighborhood, all nodes in gray perceive the central black node as the one with the lowest $\kappa_v$ in this
h-neighborhood. As a consequence, the black node also perceives itself as having the lowest $\kappa_v$ in its h-neighborhood. This
means the black node has $S_v = 1$ and therefore is the critical local node of this h-neighborhood, \textit{i.e.}, it is the node whose
removal would cause the most impact on the robustness of this particular h-neighborhood.  

It is important to remark that the proposed methodology can be implemented in a fully distributed way. 
As each node only needs local knowledge about a h-neighborhood surrounding it, 
the proposed methodology can be implemented to analyze complex undirected networks~(not necessarily only techno-social or communication networks)
and locate the fragile points of these networks in offline mode in multi-core environments with shared or distributed memory as long as there 
is full knowledge of the considered networks. In the case of P2P-like networks, such as router-level or online social networks
for instance, the proposed methodology can be implemented in parallel at each node with complexity limited to the needed local knowledge and 
thus requiring only partial and limited knowledge of the network. Dealing with directed networks (\textit{e.g.}, twitter networks) is left for 
future work.

\subsection{Navigating towards a Critical Node}
\label{subsec:nav}

Note that in practical networks one may navigate from any given node to a critical node (\textit{i.e.}, those with $S_v=1$) using the 
$\kappa_v$ and $S_v$ values assigned to each node.
This can be done following Procedure~\ref{alg:nav}. Starting from any node at the network, proceed to the node it points as having the lowest 
visible $\kappa_v$ in its h-neighborhood~(line~1), \textit{i.e.} the indicated node in the process of computing~$S_v$ for a 
h-neighborhood~(see Section~\ref{subsec:cri}).
At this point, two possible cases can arise at the indicated node: 
either (i)~it points to another node as having the lowest known $\kappa_v$ to it, \textit{i.e.}, the current node indicates another
node as the critical node it sees in its h-neighborhood; or
(ii)~it points to itself as the node with the lowest $\kappa_v$ in its h-neighborhood, \textit{i.e.} the current
node is the \emph{local} critical node for its h-neighborhood.
In case~(i), the same procedure is simply repeated until case~(ii)~occurs~(line~3). This means one can navigate through the network
following nodes with decreasing $\kappa_v$ values until reaching case~(ii). 

Upon reaching case~(ii), the current node checks its own $S_v$ score~(line~5). If $S_v = 1$,
then a critical node was reached and the navigation finishes~(line~6). If $S_v \neq 1$, there is at least one node in
the h-neighborhood seen by the current node that knows another node with a lower $\kappa_v$ than the current node.
The current node knows these nodes because they belong to its h-neighborhood, but they did not indicate it as having
the lowest  known $\kappa_v$ in their h-neighborhood. Out of these nodes that know a lower $\kappa_v$,
the current node can then randomly select one node as the next node~(line~8) and the navigation
proceeds towards it~(line~9). As this next node necessarily knows another node with a lower $\kappa_v$ than itself,
this next step in the procedure falls into~case~(i) pointing towards another node having lower~$\kappa_v$~(line~3),
ensuring that the navigation proceeds following nodes with decreasing $\kappa_v$ values until
a critical node is reached~(line~6).

Two remarks about the described navigation procedure are important. First, the navigation procedure allows no loops. This is because
the navigation procedure always progresses to nodes with lower~$\kappa_v$ than the current node or at least to a node that knows another
node with lower~$\kappa_v$ than the current node. Second, as a consequence of the first remark and as the network is finite, from
any given node one always reaches a critical node for the entire network, \textit{i.e.}, a node with $S_v=1$. 

\begin{algorithm}[t]
\footnotesize
\caption{$\textsc{navigateToCriticalNode}(currentNode)$}
\label{alg:nav}
\begin{algorithmic}[1]

\REQUIRE $currentNode$
\ENSURE $criticalNode$ 
\STATE $lowKNode \leftarrow \textsc{getLowestKNodeViewed}(currentNode)$
\IF[This is case (i)]{$lowKNode \neq currentNode$} 
	\RETURN $\textsc{navigateToCriticalNode}(lowKNode)$
\ELSE[This is case (ii)] 
	\IF{$\textsc{getSValue}(currentNode) = 1$}
		\RETURN $currentNode$ \COMMENT{Navigation finished}
	\ELSE
		\STATE $nextNode \leftarrow \textsc{getNodeKnowsLowerK}(currentNode)$
		\RETURN $\textsc{navigateToCriticalNode}(nextNode)$
	\ENDIF
\ENDIF
\end{algorithmic}
\normalsize
\end{algorithm}

\section{Experimental Results}
\label{sec:Res}

We evaluate the proposed methodology in different networks, ranging from synthetic networks to real-world ones. The idea is to study
the performance and behavior of the proposed methodology when locating the critical points of networks with different characteristics
and at different scales.

\subsection{Synthetic networks}
\label{subsec:synt-nets}

We first evaluate the proposed methodology on synthetically generated networks following the
Erd\"{o}s-R\'{e}nyi~(ER)~\cite{Erdos1959} and Barab\'{a}si-Albert~(BA)~\cite{Barabasi1999} models
for random and scale-free networks, respectively. Such networks have well-known
properties and behavior under certain conditions, thus allowing proper comparison and
analysis of the obtained results and evaluation on how the proposed methodology performs on them.

The proposed methodology intends to identify and locate nodes that are critical to network robustness.
Therefore, the rationale of the evaluation performed here is to use the
results of the proposed methodology as a strategy for identifying nodes for targeted attacks against
the considered networks and then assess the impact of this on the network robustness as compared
to classical strategic attacks, such as first removing the highest degree node. Therefore,
we use the proposed methodology to select a critical node to be removed from the original network
and then from a replica of this same network we remove the highest degree node. Next, we 
compute and compare the changes in the spectral gap $\lambda_2$ 
caused by the different attack strategies against the same network. In the cases where the
proposed methodology returns more than one critical node, just one is randomly chosen for removal.
Likewise, in the strategy of first removing the highest degree node, if there is more than one
node with the same highest degree, just one is randomly chosen for removal.

\subsubsection{Barab\'{a}si-Albert~(BA) networks}
The BA networks we use have 1,000 nodes and are generated considering 2 connections generated for each newly attached node.
This kind of scale-free networks are chosen because they are known to be particularly vulnerable to strategic attacks that
first remove their highest degree node at each step~\cite{Albert2000a} and therefore provide a suitable basis for comparison.

\begin{figure*}[ht]
 \subfigure[BA networks]{
 \includegraphics[width=0.91\columnwidth,keepaspectratio=true]{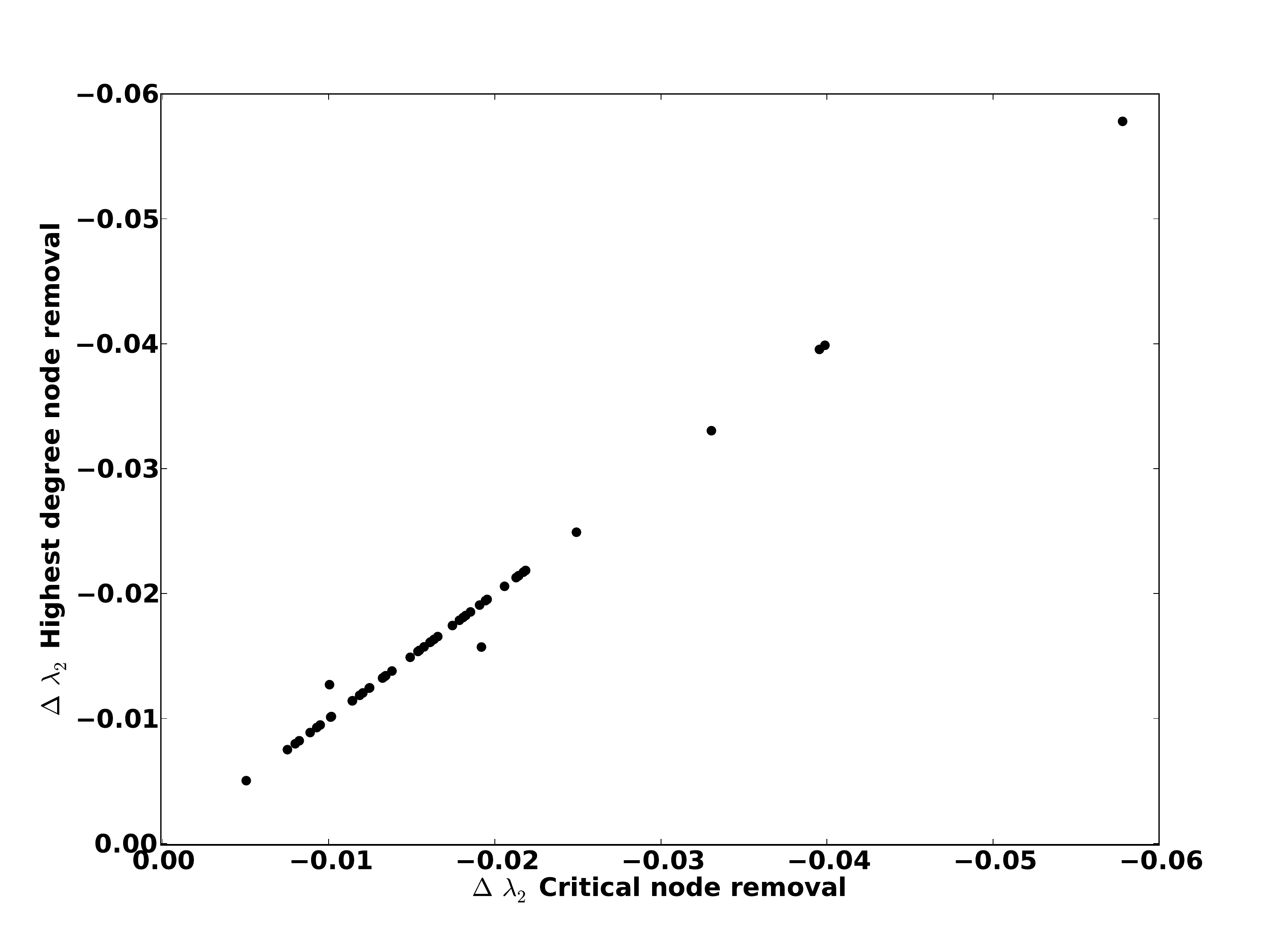} \label{fig:BACorr}}
 \subfigure[ER networks]{
 \includegraphics[width=\columnwidth,keepaspectratio=true]{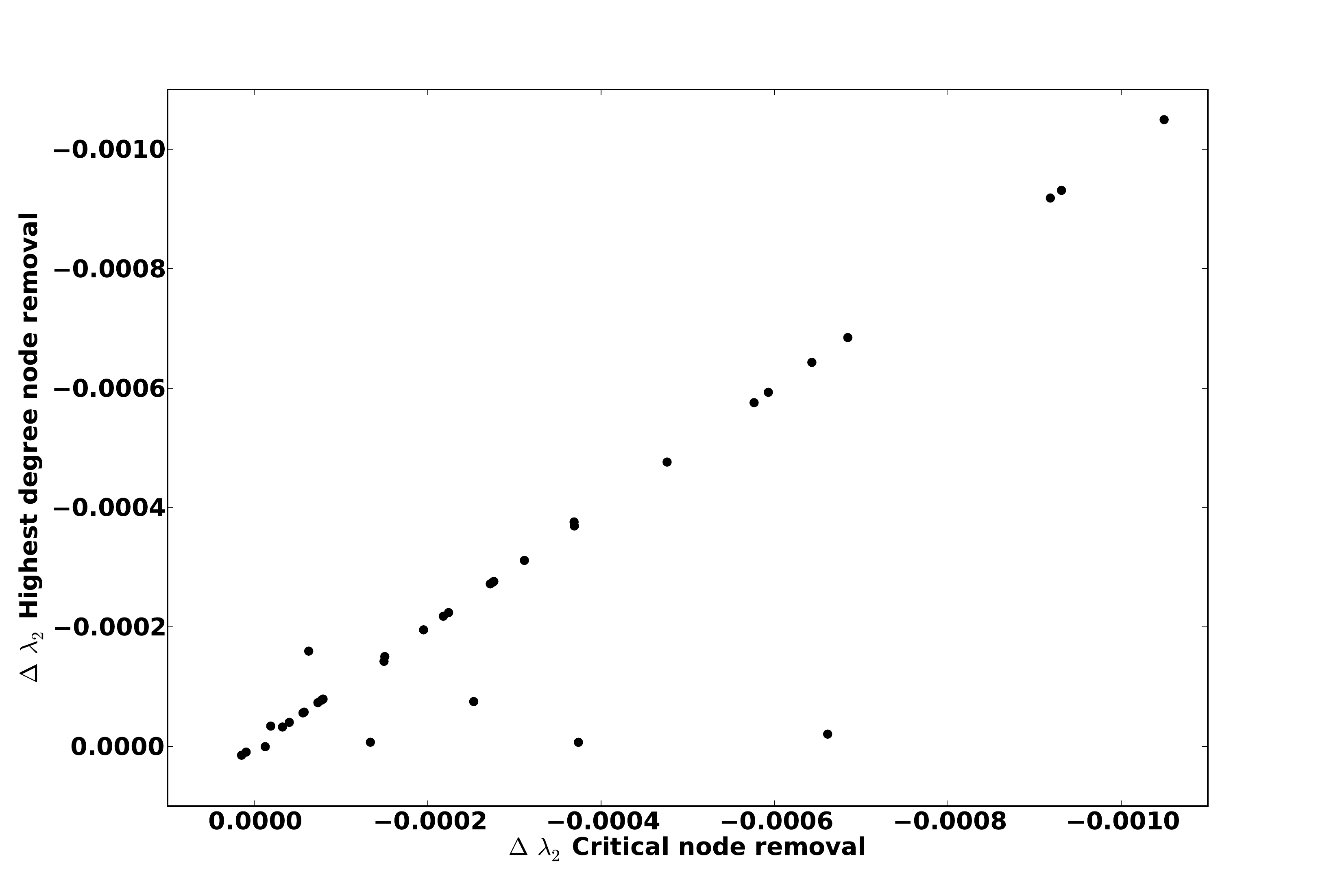} \label{fig:ERCorr}}
 \caption{Impact on the spectral gap $\lambda_2$ for both node removal strategies in BA and ER networks.}
 \label{fig:Corr}
 \end{figure*}
 

Figure~\ref{fig:BACorr} shows a scatter plot of the $\lambda_2$ variation (represented by $\Delta \lambda_2$),
\textit{i.e.}, the negative impact on the network connectivity level represented by a decrease in the spectral gap
of each network graph obtained from 50 BA~networks when both attack strategies are applied at the same network. 
The h-neighborhood used for running the proposed methodology in this experiment was empirically set to~$h=4$.
The correlation between variations in $\lambda_2$ due to both attack strategies is quite high~($R^2=0.9976$), indicating
that both strategies impact BA networks similarly. This is actually expected as BA networks are known to be vulnerable to
strategic attacks based on first removing the highest degree nodes because they provide key network connectivity.
The proposed methodology is successful in identifying such nodes as the critical nodes in the network, thus rendering 
quite similar $\lambda_2$ variations as a consequence, and hence high correlation between the strategies.
This result stresses the effectiveness of the proposed methodology in identifying the critical nodes in BA networks.

\subsubsection{Erd\"{o}s-R\'{e}nyi~(ER) networks}
The considered ER networks have 1,000 nodes and an attachment probability of 0.0045 for each node pair.
ER networks are known to be less sensitive than BA networks to the strategic attack of first removing the highest degree node~\cite{Albert2000a}.
Figure~\ref{fig:ERCorr} shows a scatter plot of the $\lambda_2$ variations obtained from 50 ER~networks when
both attack strategies are used at the same network. 
The h-neighborhood used for running the proposed methodology in this experiment was 
empirically set to~$h=6$. ER networks are known to be less vulnerable then BA networks to strategic attacks
based on first removing the highest degree node; indeed the impact in $\lambda_2$ is significantly smaller
in the results of Figure~\ref{fig:ERCorr} than in those of Figure~\ref{fig:BACorr}. Although ER networks
face a smaller degradation in network connectivity under a strategic attack of first removing the
highest degree node, this strategy also causes degradation on them and, for most cases, the highest
degree node is rather the critical node indicated by the proposed methodology. This is also indicated by a high correlation~($R^2=0.8938$)---although
not as high as in the case of BA networks---between the $\lambda_2$ variations for both attack strategies. In some cases,
the most critical node of the network---in the sense of the node whose removal causes the highest negative impact
in the level of network connectivity measured by $\lambda_2$---was clearly not the highest degree node. These are
the cases for the four dots found below the main line in Figure~\ref{fig:ERCorr} that indicate critical nodes identified
by the proposed methodology in certain networks that had a more significant negative impact in network connectivity
by their removal than the highest degree nodes in these same networks.

Note that the nodes identified by the proposed methodology typically cause the most negative impact on the network connectivity 
level measured by $\lambda_2$ in the case of their removal, even if they are not the highest degree node as in certain networks.
This leads to the conclusion that in the considered networks the methodology correctly identifies and locates
the critical node that would cause the most damage to the network connectivity in the case of its removal.

\subsection{Fragile networks}
\label{subsec:fragile}

As seen in Section~\ref{subsec:synt-nets}, there are cases where the critical node to
network connectivity is not the highest degree node. This is the case for
the (sub)network topology shown in Figure~\ref{fig:subn1}. Clearly, the critical node is the black one 
as if it is removed a major network partition would happen resulting in two large components. This kind of case represents what
we refer to as \emph{fragile} network---a network where the removal of one node causes the fragmentation of a connected network
in two or more connected components while the smaller components together represent a significant portion of the original network.
In this subsection, we analyze the capacity of the proposed methodology in locating the critical node in the
case of fragile networks.




To conduct the performance evaluation for fragile networks, 
we first generate synthetic networks that have such points of fragility. To achieve this, we start with
networks as those described in Section~\ref{subsec:synt-nets}~(both BA and ER as explained later in this subsection)
and remove nodes until we get a network that has a fragility point that fragments the network by the removal
of a single node. The node removal process up to this point in this case is conducted following a
probability for a node to be chosen for removal proportional to its degree~\cite{Wehmuth2010}.
By analyzing the traces resulting from this process, we can identify a point where the removal of a chosen 
network causes the network to fragment in a significant way. Therefore, taking the network as it was right before the removal of this node, we have
a fragile network. It is important to notice that this node whose removal fragments the network is usually not the highest degree node in the network.
As a consequence, a deterministic attack strategy of first removing the highest degree node does not select the real critical node of the network in this case.
We also remark that since the sequence of node removals that leads to the fragility point is random, although biased by the node degrees, the node found by this process is not 
necessarily the one that most severely fragments the network.
We thus expect that the removal of the critical node identified by the proposed methodology 
should partition the network in separate connected components at least as large as the ones found on the process
synthetically generating the considered fragile networks.

\subsubsection{Barab\'{a}si-Albert~(BA) fragile networks}
Conducting the experiment on 10 BA fragile networks, the results obtained show that in all cases at least one critical node is
located by the proposed methodology. In every case, the removal of these critical nodes led to a fragmentation of the network that was at least
as severe as the one resulting from the process of generating the fragile network. 

\begin{table*}[ht]
\caption{Impact of removing the critical node(s) in fragile networks.}
\label{tab:fragile}
\begin{center}
\begin{tabular}{|l|c|c|l|}
\hline
Net  & $h$ & Critical & \multicolumn{1}{c|}{Resulting Connected Components} \\ \hline
\multicolumn{4}{c}{BA fragile networks} \\ \hline
\multirow{2}{2mm}{N1} & - & 61 & 429, 7, 2, 2, 2, 2, 1, 1, 1, 1, 1, 1, 1, 1, 1 \\ 
 & 6 & 61 & 429, 7, 2, 2, 2, 2, 1, 1, 1, 1, 1, 1, 1, 1, 1 \\ \hline
\multirow{2}{2mm}{N2}  & - & 17 & 202, 19, 7, 4, 4, 3, 3, 2, 1, 1 \\ 
 & 6 & 17, 84 & 105, 63, 24, 19, 7, 4, 4, 4, 4, 3, 3, 2, 1, 1, 1 \\ \hline
\multirow{2}{2mm}{N3} & - & 22 & 432, 18, 10, 2, 1, 1, 1, 1, 1 \\ 
 & 4 & 22 & 432, 18, 10, 2, 1, 1, 1, 1, 1 \\ \hline
\multirow{2}{2mm}{N4} & - & 18 & 471, 3, 1, 1, 1, 1, 1, 1, 1 \\ 
 & 4 & 65 & 469, 4, 3, 1, 1, 1, 1, 1 \\ \hline
\multirow{2}{2mm}{N5} & - & 710 & 441, 33 \\ 
 & 6 & 6 & 403, 43, 7, 6, 4, 2, 2, 1, 1, 1, 1, 1, 1, 1 \\ \hline
\multicolumn{4}{c}{ER fragile networks} \\ \hline
\multirow{2}{2mm}{N6} & - & 484 & 170, 58 \\ 
 & 9 & 46 & 167, 60, 1 \\ \hline
\multirow{2}{2mm}{N7} & - & 966 & 142, 89 \\ 
 & 4 & 621, 173 & 140, 65, 19, 4, 1, 1 \\ \hline
\multirow{2}{2mm}{N8} & - & 88 & 57, 38, 12, 6 \\ 
 & 4 & 88 & 57, 38, 12, 6 \\ \hline
\multirow{2}{2mm}{N9} & - & 881 & 111, 43, 26, 5, 1, 1 \\ 
 & 7 & 881 & 111, 43, 26, 5, 1, 1 \\\hline 
\multirow{2}{2mm}{N10} & - & 871 & 131, 21 \\ 
 & 4 & 373, 631 & 127, 16, 3, 3, 1, 1 \\ \hline
\end{tabular}
\end{center}
\end{table*}

In order to illustrate such results, we present 
in Table~\ref{tab:fragile} a simplified result of 5~BA~analyzed networks. Each considered BA fragile network N1 to N5 is presented in two rows explained
in the following. The first row 
refers to the removal of the critical node chosen by the process of generating the fragile network. It is composed by the network id,
the critical node identified on the process of generating the fragile network, and 
the set of the distinct connected components represented by their sizes resulting from removing the chosen critical node.
The second row refers to the removal of the critical node identified by the proposed methodology. It consists of the $h$ values used by the 
proposed methodology to determine the h-neighborhoods to be considered around each node, the critical nodes located by the proposed methodology,
and the size of the connected components resulting from removing these critical nodes. In the case where more than one critical
node is located by the proposed methodology, they are all removed.

Analyzing the results presented in Table~\ref{tab:fragile}, we observe that for networks N1 and~N3 both methods 
indicate the same nodes as critical to network connectivity. As a consequence, after removing these critical nodes,
the resulting connected components in these cases are the same. In the cases for networks N2 and N5, the proposed
methodology locates critical nodes that fragment the network in a larger number of resulting components as compared
to the removal of the critical node chosen by the process of generating the fragile network. Note that, for the fragmentation
caused by the removal of the critical nodes located by the proposed methodology, the fragmented
resulting connected components (\textit{i.e.}, excluding the main largest resulting component) in these cases are
also typically larger in size. This characterizes a stronger fragmentation of the network. For example, in the case
of network N5, removing the critical node~(\#6) located by the proposed methodology fragments the considered network 
composed by 475 nodes into 14 different components, effectively disconnecting 71 nodes from the network. In contrast,
removing node \#710 only fragments the network into 2~resulting components while excluding only 33 nodes.
Clearly, the critical node pointed out by the proposed methodology has a higher impact in the network connectivity.
Note that for network N2, the proposed methodology located an additional critical node~(\#84) besides the critical node~\#17.
This allowed a fragmentation of the network significantly higher when considering the critical nodes pointed out
by the proposed methodology.
Network~N4 also offers an interesting case: removing the critical node~(\#65) located by the proposed methodology actually
fragments the network into fewer resulting connected components. Nevertheless, the impact on the 
network connectivity is still higher as more nodes are disconnected in this way~(12 nodes against~10).

\subsubsection{Erd\"{o}s-R\'{e}nyi~(ER) fragile networks}

A similar experiment is then repeated for 10 ER fragile networks. Table~\ref{tab:fragile} presents a simplified 
result of 5~ER~analyzed networks (identified as N6 to N10). 
The analysis of the results achieved by the proposed methodology for ER fragile networks
in Table~\ref{tab:fragile} is similar to the one performed for the BA fragile networks. Likewise, the conclusions
also suggest that the critical nodes pointed out by proposed methodology achieve a fragmentation at least
equivalent~(as in the cases of network N8 and~N9), if not stronger~(as in the cases of networks N6, N7, and~N10) than the
reference for comparison.


\subsection{Real-world network trace}

In this subsection, we evaluate the proposed methodology in locating critical nodes using a real-world network trace.
The connected network extracted from this real-world trace is composed of 190,914 nodes representing a router-level
network topology collected by CAIDA.\footnote{\url{http://www.caida.org/tools/measurement/skitter/router_topology/}}
This network has a diameter of~26 as well as an average and maximum node degrees of 6.34 and 1071, respectively.
Executing the proposed methodology using an experimentally set $h=4$ indicates node~40412 as the single most critical node 
for the whole network. The 
removal of this single critical node fragments the network into three relatively large connected components having
189608, 1184, and~121 nodes, characterizing a major disruption in the network.

The main point of evaluating the proposed methodology for this real-world network trace is rather checking out
the feasibility of applying it on a large scale network. Despite having over 190 thousand nodes, our proposed
methodology can locate the critical nodes considering only localized information of the h-neighborhoods
around each node (here with $h=4$) in a fully distributed way. 
The smallest and the largest considered h-neighborhoods have 5 and 123451 nodes, respectively. 
Moreover, the average size of the considered h-neighborhoods
is 9023 nodes, thus limiting the complexity of computing~$\kappa_v$ for each node. 

%
%


%
%
%
%
%

\section{Related Work}
\label{sec:RelW}

Network robustness is an important property derived from the connectivity level
that directly impacts network reliability. There are many studies investigating
network robustness in general and methods to evaluate network connectivity level~\cite{Albert2000a,Newman2000,Kim2004}.
Nevertheless, to the best of our knowledge, only a few recent works target the distributed evaluation and location of the most critical nodes to network robustness, thus assessing node centrality~\cite{Nanda2008,Kermarrec2011,Dinh2010} in a distributed way.

Nanda and Kotz~\cite{Nanda2008} propose a new centrality metric called 
Localized Bridging Centrality (LBC). LBC is evaluated using only one hop neighborhood around 
each node for which it is calculated. The proposed use of this method is on relatively small scale wireless mesh networks. It can 
to a certain extent perceived as a specialization of the general method we propose, restricting the h-neighborhood to $h=1$.
One of the main motivations for the work of Nanda and Kotz 
was a paper published in 2002 by Marsden~\cite{Marsden2002}, which shows empirical evidence 
that localized centrality measures calculated for one hop radius neighborhood are highly correlated to the global centrality measure.  
Kermarrec~et~al.~\cite{Kermarrec2011} propose a new centrality measure, called second order centrality.
The second order centrality is defined in terms of the standard deviation of the time between visits of a perpetual random walk to each node.
This method has the same goal as ours in identifying the critical nodes in the network in a distributed way without requiring full knowledge 
network topology. Nevertheless, relying on perpetual random walks has a potentially long and indeterminated convergence 
time, while our approach offers a faster and deterministic convergence time.
Both Jorgi\'{c}~et~al.~\cite{Jorgic2004} and Sheng~and~Li~\cite{Sheng2006} propose localized and distributed methods for for critical 
node detection in mobile ad hoc networks. These methods are specific for wireless ad hoc networks and use both topological and
spacial properties of the network.
Dinh~et~al.~\cite{Dinh2010} propose a new model to assess network vulnerabilities formulating it as an optimization problem
that can render approximate solutions with provable performance bounding. This method uses full knowledge of network topology, hindering its
applicability to large scale networks where such an information may not be available and distributed implementation is required.

Spectral analysis~\cite{Chung1997} has been previously used for network analysis in different ways.
Jamakovic and Uhlig~\cite{Jamakovic2007a} study the relation between the algebraic connectivity~\cite{Fiedler1973}and the network robustness to node and link failures.
The algebraic connectivity, \textit{i.e.} the second smallest eigenvalue of the Laplacian 
matrix, is a spectral property of a graph, which is an important parameter in the analysis of various robustness-related problems. Network robustness is quantified with the node and the link connectivity, two topological metrics that give the number of nodes and links 
that have to be removed in order to disconnect a graph.
The conclusion reached is that there is a non trivial relation between algebraic connectivity and node/link connectivity.
The algebraic connectivity is thus equivalent to the notion of spectral gap adopted in our work. Bigdeli~et~al.~\cite{Bigdeli2009} report on an effort to compare different network topologies according to their algebraic connectivity, network 
criticality, average node degree, and average node betweenness, showing that each one of them captures different characteristics of the 
network and is better suited to a certain class of problems. Gkantsidis~et~al.~\cite{Gkantsidis2003} analyze
the Internet structure using spectral analysis. Based on spectral analysis, Fay~et~al.~\cite{Fay2010} have recently derived a new metric called 
weighted spectral distribution to perform structural comparison of different networks. 
These efforts concentrate their analysis on the relation between network robustness and the spectral gap (or related and derived metrics) for the whole network, while we use the notion of \emph{localized} spectral gap~(limited to a given neighborhood of each node) to perform
a distributed assessment of network centrality as well as location of the critical nodes.

\section{Summary and Outlook}
\label{sec:Conc}

We propose a localized and distributed methodology capable of locating critical nodes 
to network robustness based on the analysis of the spectral gap of a h-neighborhood
around each node. Such critical nodes may be thought of as those the most related to the notion of network centrality.
The proposed methodology is shown to be well suited
for distributed implementation and it does not require knowledge of the full network
topology. We also propose a procedure that allows one to
navigate from any node towards a critical node following
only local information computed by the proposed methodology.
Results obtained for different kinds of networks and at different scales
confirm the effectiveness of the proposed methodology in
locating the most critical nodes to network robustness.

The encouraging results obtained using the proposed methodology lead to interesting
perspectives for future work:

\begin{itemize}

\item h-neighborhood determination -- during the experimental evaluation of the proposed methodology we could observe that the adopted 
radius to build the h-neighborhoods around each node influences the obtained results. 
On the one hand, If the considered $h$ is too small, the resulting h-neighborhoods can also be relatively small
and might lead to critical nodes of restricted local concern, thereby not representing the topological features
in terms of robustness of the whole network. On the other hand, if $h$ is set too large, the resulting h-neighborhoods 
can also be too large, rendering excessively high the computational cost of determining the local value $\kappa_v$ for each node 
$v$ to still point out the same critical nodes that would be found using a smaller radius. Yet larger $h$ values may
eventually lead to many or all h-neighborhoods in fact comprising the whole network, thus degrading the methodology
since all these would yield the same value for their local $\kappa_v$ values.
For the presented results in this paper, the most suitable $h$ value for each network has been set experimentally. 
To overcome this limitation, we intend as future work to address the issue of defining a solution to (at least approximatively) 
determining the most suitable $h$ to be adopted in each network case. 

\item Other metrics to assess local robustness -- in this paper, the local $\kappa_v$ value representing the relative importance of 
each node $v$ to the (local) network robustness is computed based on the spectral gap of a h-neighborhood around each
node~$v$. However, we understand that other metrics besides the spectral gap may be considered to 
determine $\kappa_v$. This would lead to different alternative ways of determining $\kappa_v$, 
eventually being less costly than using the spectral gap or generating better results.
Further, we may possibly use alternative metrics that are specific to certain kinds of networks (\textit{e.g.}, ad-hoc 
wireless networks), lending to better results for particular networks. Hence, we intend to investigate
alternative metrics to determine the local $\kappa_v$ value at each node.

\item Network partitioning technique -- considering the proposed navigation procedure, each node in the network is
associated to one and only one critical node. This 1-to-1 association relationship 
may be thought of as creating a \emph{network partition} where each critical node determines an equivalence class.
Exploring the possibility of using the proposed methodology as a network partitioning technique and studying the
properties of the resulting network partitions is left for future work.

\end{itemize} 
%
%



\section*{Acknowledgements}

This work was partially supported by the Brazilian Funding Agencies FAPERJ, CNPq, CAPES, and by the Brazilian Ministry of Science and Technology (MCT).
Authors thank \'Eric Fleury (ENS-Lyon/INRIA, France) for comments on an earlier version of this paper.



\end{document}